%% file: prl_05072003.tex
\documentstyle[prd,floats,aps,epsfig,preprint]{revtex}
\newcommand{\whichformat}{}

\newlength{\pushupfigure}
\def \epsfin_v1#1#2{
        \vspace{\pushupfigure}
        \center
        \leavevmode 
        \epsfxsize=#1
\newcommand{ \qbar}     {\mbox{$\overline{q}$}}
        \epsffile[20 143 575.75 698.75]{#2}
}
\newcommand{\gev}  {\mbox{${\rm GeV}$}}

\newcommand{\gevc} {\mbox{${\rm GeV}/c$}}
\newcommand{\gevcc}{\mbox{${\rm GeV}/c^2$}}

\newcommand{\ipb}{\mbox{${\rm pb}^{-1}$}}

\newcommand{\pt}{\mbox{$p_{T}$}}
\newcommand{\et}{\mbox{$E_{T}$}}
\newcommand{\met}{\mbox{${E\!\!\!\!/_{T}}$}}
\newcommand{\metvec}{\mbox{${\vec{E}\!\!\!\!/_{T}}$}}

\newcommand{\phiellmet}{\mbox{$\phi_{\ell E\!\!\!\!/_{T}}$}}
\newcommand{\ptl}{\mbox{$p_T^{\ell}$}}
\newcommand{\Set}{\mbox{$\Sigma p_T (\ell ,\tau_{h} , \met )$}}
\newcommand{\pttauh}{\mbox{$p_T^{\tau_h}$}}
\newcommand{\degr}{\mbox{$^{\circ}$}}
\newcommand{\ppbar}{\mbox{$p\overline{p}$}}
\newcommand{\qqbar}{\mbox{$q\overline{q}$}}

\newcommand{\ttbar}{\mbox{$t\overline{t}$}}
\newcommand{ \qbar}     {\mbox{$\overline{q}$}}

\newcommand{ \Wjet}{\mbox{$W$+jets}}
\newcommand{ \Wp}{\mbox{$W^+$}}
\newcommand{ \Wm}{\mbox{$W^-$}}
\newcommand{ \Wpm}{\mbox{$W^\pm$}}
\newcommand{ \Zz}{\mbox{$Z$}}
\newcommand{ \Zg}{\mbox{$Z/\gamma^*$}}
\newcommand{ \Zll}{\mbox{$\Zz \to \ell^+ \ell^-$}}
\newcommand{ \Ztautau}{\mbox{$\Zz \to \tau^+ \tau^-$}}
\newcommand{ \Zgtautau}{\mbox{$\Zg \to \tau^+ \tau^-$}}
\newcommand{ \Zptautau}{\mbox{$\Zz (\to \tau^+ \tau^-)$}}

\newcommand{\tauh}{\mbox{$\tau_{h}$}}
\newcommand{\rpv}{\mbox{${R\!\!\!\!\!\:/_p}$}}
\newcommand{\rp}{\mbox{${R_p}$}}
\newcommand{\lampthree}{\mbox{$\lambda_{333}'$}}

\newcommand{ \stopone}  {\mbox{$\tilde{t}_{1}$}}
\newcommand{ \stoponen}  {\tilde{t}_{1}}
\newcommand{ \stoponeb} {\mbox{$\bar{\tilde{t}}_{1}$}} 
\newcommand{ \stoppair}{\stopone\stoponeb}
\newcommand{ \mystop}{\mbox{stop}}

\newcommand{ \mtlmet}{\mbox{$M_T (\ell , E\!\!\!\!/_{T} )$}}
\newcommand{\etal}{{\it et al.}}
\newcommand{\ISAJET}{{\sc isajet}}

\newcommand{\VECBOS}{{\sc vecbos}}
\newcommand{\HERWIG}{{\sc herwig}}

\newcommand{\br}{{\rm Br}}
\newcommand{\finallimit}{$122$}

\newcommand{\mytableorig}{
\begin{table}
\whichformat
\begin{minipage}[b]{7.05in}
\begin{tabular}{cccccc|cc}
Sample			   & $t \bar{t}$   & Diboson      &$W+{\rm jets}$    &$\Zgtautau$  & QCD        & Tot  	& N$_{\rm Obs}$  \\ \hline\hline
 OS $\ell\tau$		   &1.2$\pm$0.3    &2.3$\pm$0.8   &101$\pm$6    &225$\pm$9   & 301$\pm$18 & 631$\pm$21	& 642  \\ \hline
$\ell\tau_{h} +\ge 2$ jets &1.0$\pm$0.2	   &0.4$\pm$0.1   &3.4$\pm$0.4 &7.7$\pm$0.5 & 8$\pm$3    & 21$\pm$3 	& 16  \\ \hline
$\mtlmet< 35$~GeV/$c^2$	   &0.15$\pm$0.07  &0.14$\pm$0.06 &0.5$\pm$0.2 &6.0$\pm$0.4 & 8$\pm$3    & 15$\pm$3	& 10  \\ \hline
$\Set > 75$~\gevc		   &0.15$\pm$0.07  &0.08$\pm$0.03 &0.2$\pm$0.1 &2.8$\pm$0.3 & $0^{+1.4}_{-0}$   & 3.2$^{+1.4}_{-0.3}$ 	& 0   \\
\end{tabular}
\caption{Summary of the number of OS events in the data and expectations for the background sources as each selection requirement is applied.}
\label{tab:DataReduction}
\end{minipage}
\end{table}

}

\newcommand{\myfigIII}{
\begin{figure}[t]
\vspace*{30.mm}
\begin{center}
\epsfig{file=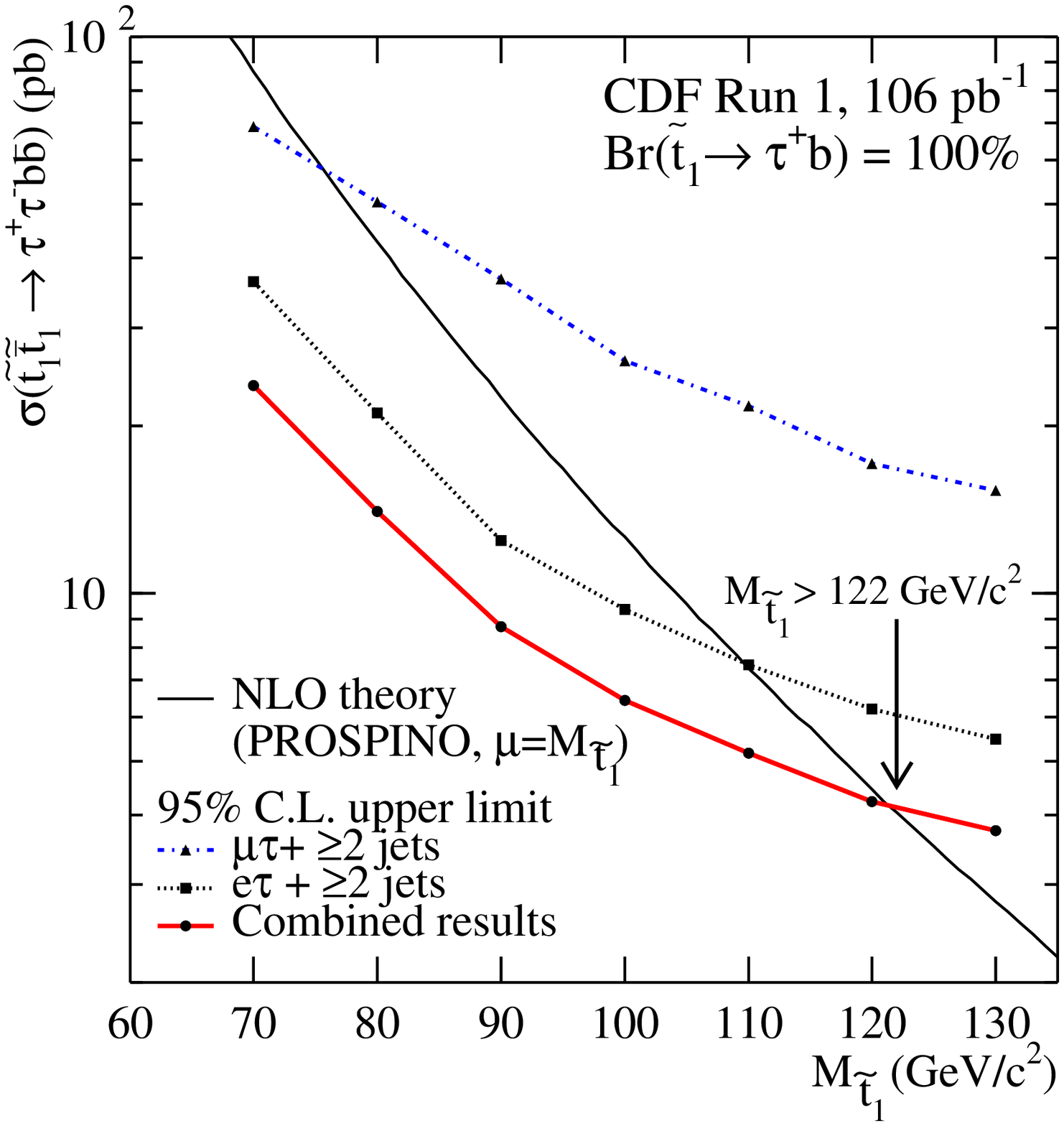,width=250pt}
\caption{The 95\% C.L. upper limit on cross section 
for  $\stopone\stoponeb$ production
compared to the NLO calculations}
\label{fig:t1limit}
\end{center}
\end{figure}
}

\begin{document}
\begin{flushright}
\preprint{\today}

\end{flushright}


\begin{center}
\begin{bf}
 {Search for Pair Production of Scalar Top Quarks in {\it R}-parity Violating Decay Modes in \ppbar\ Collisions at $\sqrt{s} = 1.8$ TeV}       
\end{bf}
\end{center}
\begin{small}

\input{cdf_auth_030501.tex}

\end{small}
\vspace{1cm}


%
%


\begin{abstract}
\par\noindent
We present the results of a search for pair production of scalar top quarks (\stopone) in an {\it R}-parity violating supersymmetry scenario in 106 \ipb\ of \ppbar\ collisions at $\sqrt{s} = 1.8$ TeV collected by the Collider Detector at Fermilab. In this mode each $\stopone$ decays into a $\tau$ lepton and a $b$ quark. We search for events with two $\tau$'s, one decaying leptonically ($e$ or $\mu$) and one decaying hadronically, and two jets. No candidate events pass our  final selection criteria. We set a 95\% confidence level lower limit on the $\stopone$ mass  at \finallimit\ $\gevcc$ for $\br(\stopone\rightarrow \tau b) = 1$.

\end{abstract} 
\maketitle

\vspace{0.3cm}

%
%
%
\clearpage
%

Many supersymmetry (SUSY) models~\cite{mssm} predict that the first two generations of supersymmetric partners of the quarks and the leptons (squarks and sleptons) are approximately mass degenerate. However, the mass of the lightest
top squark (\stopone\ or `stop') can be relatively light due to a
large mixing between the interaction eigenstates, $\tilde{t}_L$ and
$\tilde{t}_R$.  This mixing depends in part on the top Yukawa coupling
which is largely due to the heavy top quark mass, and it is possible
that \stopone\ is lighter than the top quark~\cite{lstop}.

%
%

$R$-parity ($\rp$) is a multiplicative quantum number defined as $\rp \equiv (-1)^{3B+L+2S}$, where $S$, $B$ and $L$ are the spin, baryon and lepton numbers of a particle, respectively~\cite{rparity}. $\rp$ distinguishes SM particles ($\rp = +1$) from SUSY  particles ($\rp = -1$).  Conservation of $\rp$ requires SUSY particles to be produced in pairs and to decay, through a cascade, to SM particles plus the stable lightest supersymmetric particle. The $\rp$ conservation, which is not required by SUSY, is often built into the theory by hand and is justified phenomenologically by limits on the proton lifetime, the absence of flavor-changing neutral currents, etc. Viable $\rp$ violating (\rpv) models can be built by adding explicit $\rpv$ terms with trilinear couplings  ($\lambda_{ijk}$, $\lambda_{ijk}^{\prime}$, $\lambda_{ijk}^{\prime\prime}$) and spontaneous $\rpv$  terms with bilinear couplings ($\epsilon_{i}$) to the SUSY Lagrangian~\cite{rpvtheory1,rpvtheory2}, where $i$, $j$ and $k$ are the generation indices. These couplings allow  $B$ or $L$ violating interactions and, if $\lambda^\prime_{33k}$ or $\epsilon_{3}$ is non-zero, a \stopone\ may decay directly to SM final states which are experimentally observable.


%
%
At the Fermilab Tevatron, in $p\bar{p}$ collisions, stop pairs might be produced strongly via \rp-conserving processes through $gg$ fusion and $\qqbar$ annihilation. In $\rpv$ scenarios each stop can decay into a tau ($\tau$) lepton and a bottom ($b$) quark with a branching ratio, $\br$, which depends on the coupling constants of the particular model. A good final state search topology identifies either an electron or a muon ($\ell$ = $e$ or $\mu$) from the $\tau \rightarrow \ell \nu_{\ell} \nu_{\tau}$ decay, as well as a hadronically decaying tau ($\tau_h$) lepton, and two or more jets. 

%
%
We present the results of a search for $\stopone\stoponeb \rightarrow \ell\tau_h jj$ events, in the framework of \rpv-MSSM, using 106 \ipb\ of $\ppbar$ collisions at $\sqrt{s}$ = 1.8 TeV collected by the Collider Detector at Fermilab (CDF) during the 1992$-$95 run of the Tevatron (Run~I). CDF is a general purpose detector and has been described in detail elsewhere~\cite{CDFdet,Topref}. We briefly describe the subsystems of the CDF detector relevant to this analysis. The location of the \ppbar\ collision event vertex ($z_{vtx}$) is measured along the beam direction~\cite{cdfdefs} with a time projection chamber. The transverse momentum ($\pt$) of charged particles is measured in the region $|\eta|<1.0$  by a central tracking chamber (CTC) which is immersed in a uniform 1.4~T solenoidal magnetic field~\cite{cdfdefs}. Electromagnetic (EM) and hadronic (HAD) calorimeters, segmented in a projective tower geometry surrounding the solenoid and covering the region $|\eta|<4.2$, are used for identification of electrons, taus, and jets and the measurement of the missing transverse energy ($\met$). The central strip chamber (CES) is embedded in the central EM calorimeter at a depth of approximately shower maximum, and is used for further electron identification as well as $\pi^0 \rightarrow \gamma\gamma$ identification from $\tauh$ decays. A muon subsystem is located outside the hadron calorimeter and has trigger coverage for the region $|\eta|< 0.6$.

%
%
%

The analysis begins with a sample of events which pass a three-level trigger system~\cite{CDFdet} which requires a single isolated lepton ($e$ or $\mu$) with $\pt >$~8~GeV/$c$ ($|\eta|<1.0$ if it is an electron and $|\eta|<0.6$ if it is a muon)~\cite{lowpttrig}. Offline, the lepton is required to have \pt~$>10$~GeV/$c$, come from the event vertex, and pass more restrictive identification and isolation requirements~\cite{Topref,IsoReq}. An event is removed as a $\Zz$ boson candidate if it contains a second, loosely identified same-flavor opposite-sign lepton with $76 < M_{\ell\ell} <106$~\gevcc. To maintain the projective geometry of the calorimeter, all events are required to have $|z_{vtx}| \le 60$ cm.

%
%
An inclusive $\ell\tau_h$ subsample is made by requiring each event to further contain a high \pt, isolated, hadronically decaying $\tau$ lepton candidate with $\pttauh > 15\ \gevc$~\cite{ETDef} and $|\eta|<1.0$. A $\tau_h$ candidate is identified as a calorimeter cluster which satisfies the following requirements~\cite{pub:tauh}:  
(i)~not identified as an $e$ or $\mu$; 
(ii)~one or three tracks with $\pt > 1\ \gevc$ in a 10\degr\ cone around
the calorimeter cluster center;
(iii)~the scalar sum of the \pt\ of all tracks in $\Delta R = 0.4$ around the cluster center,
excluding those in the $10\degr$ cone,  less than 1 \gevc;
(iv)~fewer than three $\pi^0 \to \gamma\gamma$ candidates identified in  the CES;
(v)~more than $4\ \gev$ of $\et$ measured in the calorimeter;
(vi)~$0.5 < \et/\pttauh < 2.0\ (1.5)$ for one track (three tracks);
(vii) the cluster width 
of the calorimeter cluster in $\eta$-$\phi$ space
 less than $0.11 \; (0.13) - 0.025 \; (0.034) \times E_{T} \; [\gev]/ 100$ for one track (three tracks); and 
(viii) the invariant mass reconstructed from tracks and $\pi^0$'s  less than 1.8 \gevcc.
The charge of the \tauh\ object is defined as the sum of the track charges, and is required to have unit magnitude and have the opposite sign (OS) of the $\ell$ candidate. A total of 642 events pass the above requirements; 16 of these have two or more jets (identified using a fixed cone algorithm with $\Delta R=0.4$~\cite{jet_reco}) with $\et > 15\ \gev$ and $|\eta|< 2.4$. Note that, as expected, the four $\ell\tau_h+$jets candidate events which were found in the search for $\ttbar \to (W^+b)(W^-{\bar b})$~\cite{pub:tauh} pass the kinematic requirements for this search.

%
%

The dominant backgrounds come from $\Zz/\gamma^*(\to \tau^+\tau^-$)+jets, $\ttbar$, diboson ($\Wp\Wm$, $\Wpm\Zz$, $\Zz\Zz$) production, and fake $\ell\tauh$ combinations from \Wjet\ and QCD events. Monte Carlo (MC) programs with CTEQ4L parton distribution functions (PDFs)~\cite{CTEQ4} and a detector simulation are used to estimate the background rates by  simulating the kinematics of $\Zz/\gamma^*,~W,~\ttbar$, and diboson events. All SM processes except $W/Z$+jets events are generated using \ISAJET~\cite{isajet}; \VECBOS~\cite{vecbos} is used for vector boson plus jets production and decay, followed by \HERWIG~\cite{herwig} for the fragmentation and hadronization of the quarks and gluons. The cross sections for $\Zz$/$\gamma^{\ast}$, $\ttbar$ and $\Wp\Wm$ production are normalized to the CDF  measurements~\cite{Z_CDF,DY_CDF,TOP_CDF,WW_CDF} and next-to-leading order (NLO) calculations for $\Wpm\Zz$ and $\Zz\Zz$ production are used~\cite{WZ_NLO,ZZ_NLO}. The number of fake events from QCD is estimated from the data and assumes that the number of OS events, after subtracting off the non-fake contribution, is identical to the number of like-sign (LS) events observed in the data as expected from QCD sources i.e., $N^{OS}_{QCD} = N_{data}^{LS} - N_{MC}^{LS}$.

%
%

\begin{figure}
\vspace{-10mm}
\begin{center}
\epsfig{file=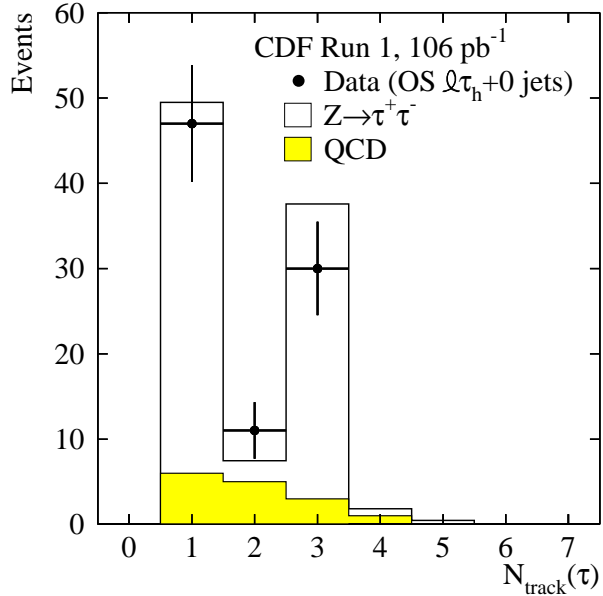,width=250pt}
\caption{The number of charged tracks in each $\tauh$ candidate for the opposite-sign (OS) $\ell\tauh$+0 jet control sample. The data are compared to the MC expectation (all background histograms are summed) which is dominated by real $\tau_{h}$'s from $\Zz \rightarrow \tau^{+}  \tau^{-}$ production.}
\label{fig:tautrack}
\end{center}
\end{figure}

The final optimized data selection requirements are based on simulated $\stopone\stoponeb$ production, using \ISAJET~\cite{isajet} and the CDF detector simulation, background expectations, and a control sample. See Fig.~2. To reduce the number of \Wjet\  events we require $\mtlmet <  35\ \gevcc$ where $\mtlmet$ is the transverse mass of the $\ell$ and the event \met, defined as $\mtlmet \equiv \sqrt{ 2\ \ptl \met (1 - \cos \phiellmet) }$, and where \phiellmet\ is the azimuthal angle difference  between the $\ell$ and the \met. To reduce the QCD backgrounds we require  $\Set \equiv \ptl + \pttauh + \met\ > 75\ \gevc$. A control sample of $\ell\tau_h$+0 jet events with similar kinematic requirements ($\mtlmet<25$~GeV/c$^2$, $|\vec{p}_{T}^{\ell} + \metvec | > 25$~GeV/$c$) is selected to show that the backgrounds are well modeled, dominated by real $\Ztautau$ production, and for later use in the acceptance calculations. Figure~\ref{fig:tautrack} shows the charged track multiplicity of the hadronic tau decays (removing the 1 and 3-prong requirements) for this sample and shows good agreement with background expectations.

\begin{figure}
\vspace*{-10mm}
\begin{center}
\epsfig{file=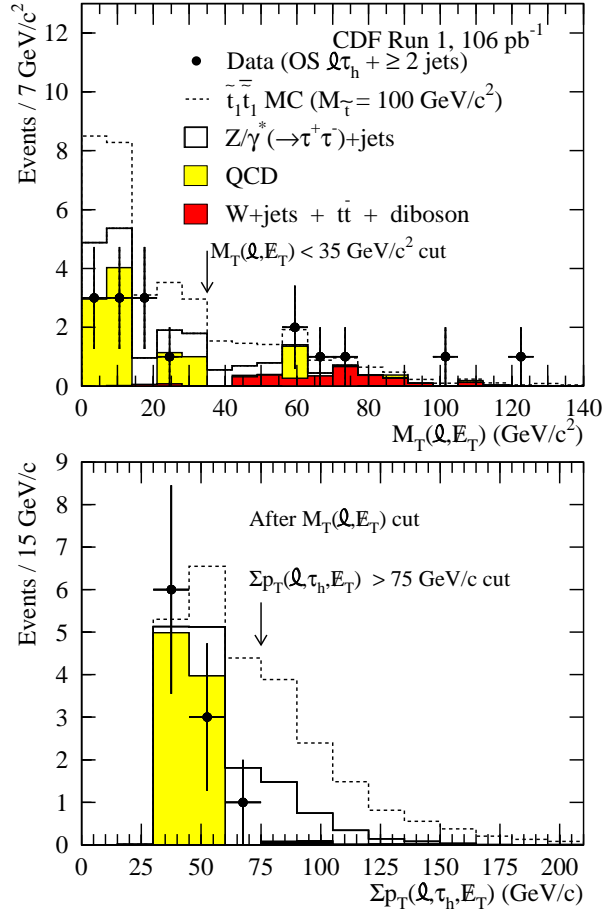,width=250pt}
\caption{The final data selection criteria for the OS $\ell\tau_{h} + \ge 2$~jet sample. The arrows show the final event selection requirements. The assumed stop mass is 100~GeV/c$^2$. The quantities $\Set$ and $\mtlmet$ are defined in the text.}
\label{fig:kinematics}
\end{center}
\end{figure}

%
%
A comparison of the OS $\ell\tau_{h} + \ge 2$~jet data and background estimation is shown in Fig.~\ref{fig:kinematics} before the final $\mtlmet$ and $\Set$ cuts. A breakdown of the backgrounds and data is given in Table~\ref{tab:DataReduction}. The backgrounds appear well modeled. A total of 3.2$^{+1.4}_{-0.3}$ events are predicted from all SM sources, dominated by $\Zptautau$+jets production. No candidate events pass the final \stoppair\ selection criteria, which is unusual but expected in roughly 3\% of experiments when taking into account the statistical and systematic uncertainties.

%
%

\myfigIII

In order to set limits on \stopone\stoponeb\ production and decay, the acceptances and efficiencies are normalized to the rate of $\Zptautau$+0 jet decays using the following relation: 

\begin{equation}
\sigma(\stoppair \rightarrow \tau^+\tau^-b\bar{b})
= \left( \frac{ N^{Obs}_{\mystop} - N^{BG}_{\mystop}
 }{ N^{Obs}_{\Zz} - N^{BG}_{\Zz} } \right) \cdot
 R_{Acc} \cdot R_{Trig} \cdot \nonumber 
\sigma_{\Zz} \cdot \br(\Ztautau)
\end{equation}
\label{eq:xstop2}


\noindent where $N_{\mystop}^{Obs}$ and $N_{\mystop}^{BG}$ 
($N^{Obs}_{\Zz}$ and $N^{BG}_{\Zz}$)
are the number of candidate events observed in the data 
and expected background in the $\ge 2$~jet/\stoppair\ (0~jet/$\Zz$) selections, $R_{Acc}$ is the ratio of the $\Zz$ to \stoppair\ acceptances and $R_{Trig}$ is the ratio of the trigger efficiencies. The primary
advantage of this approach is that potential systematic uncertainties
in the estimate of identification and isolation efficiencies are reduced in the ratio of \stopone\stoponeb\ to $\Zz$ production.


The 95\% confidence level (C.L.) limits on $\sigma(\stoppair \rightarrow \tau^+\tau^-b\bar{b})$ in the $e$, $\mu$ and combined channels are found using Eq.~(1) and come from a Bayesian integration of the likelihood as a function of the cross section, integrating over the correlated and uncorrelated systematic uncertainties on the expected signal with a flat prior. The $R_{Acc}$ term is a function of the $\stopone$ mass and varies in the range 0.34 $<R_{Acc}^e<$ 2.15 (0.35 $<R_{Acc}^{\mu}<$  1.87) for the $e$ ($\mu$) channel over the range 70 $< m_{\stoponen} < $ 130 GeV/$c^2$.  The $R_{Trig}$ term varies between $0.95<R_{Trig}^e<0.97$ $(0.99 <R_{Trig}^{\mu}<1.00$) for the $e$ ($\mu$) channel with an uncertainty of about 1\%. (The acceptance and trigger efficiencies for the $Z$ control sample for this analysis are 1.19\% (0.69\%) and 74.5\% (83.0\%) for the $e~(\mu)$ channel respectively.) Assuming lepton universality gives $\sigma_{\Zz} \cdot \br(\Ztautau)$ = $\sigma_{\Zz} \cdot \br(\Zll)$ = $231 \pm  12$ (stat+sys)~pb~\cite{pub:Zpro}. The dominant uncertainty is due to the statistical uncertainty in $N^{Obs}_{\Zz} - N^{BG}_{\Zz}$ and is 17.0\% (24.9\%)~\cite{Zuncert}. Additional uncertainty comes from our estimation of $R_{Acc}$ which is dominated by the variation in the $\stopone\stoponeb$ acceptance from choices of the QCD renormalization scale $Q^2$, PDFs,  amount of gluon radiation, the jet energy scale and the statistical uncertainty in the MC samples~\cite{IndivResults}. The total uncorrelated uncertainties vary between 17.1 and 17.7\% (25.1\% and 25.4\%), and the total correlated uncertainties vary between 9.3 and 14.1\%.

%
%

Figure~\ref{fig:t1limit} shows the final 95\% C.L. upper limits on the cross section times $\br$ for the $e$,  $\mu$ and combined channels, along with the NLO prediction of the production cross sections~\cite{PROSPINO}.  The 95\% C.L. lower limits on  $M_{\stopone}$ are 110 and 75 \gevcc\ for the $e$ and $\mu$ channels, respectively, where we have assumed $\br$ = 1 for simplicity. Combining the two results yields a limit of \finallimit\ \gevcc. Since our analysis does not distinguish the quark flavors in jet reconstruction, these results are equally valid for any $\lambda^\prime_{33k}$ coupling.  These results substantially improve on the currently most stringent mass limit which comes from the ALEPH experiment~\cite{aleph_RPVstop} which excludes $\stopone$ masses below 93 \gevcc\ using $e^+e^- \to \stopone \stoponeb \to \tau^+ \tau^- +$ 2~jets topology with an assumption of $\lambda^{\prime}_{33k} \not  = 0$ ($k$ = 1, 2 or 3).

%
%

In conclusion, we have searched for $\stopone \stoponeb$ production  using  106~$\ipb$ data in \ppbar\ collisions  at $\sqrt{s}$ = 1.8 TeV. We have examined the $\ell\tauh+\ge2$~jet final state within an \rpv\ SUSY scenario in which each  $\stopone$ decays to a $\tau$ lepton and a $b$ quark  via non-zero  $\lampthree$ or $\epsilon_{3}$ couplings. No $\stopone \stoponeb$ event candidates pass our selection criteria and we have set a 95\% C.L. lower limit on the $\stopone$ mass  at \finallimit~$\gevcc$ for $\br = 1$.


%
%

We thank the Fermilab staff and the technical staffs 
of the participating institutions for their vital contributions.  
This work was supported by the U.S. Department of Energy and
National Science Foundation;
the Italian Istituto Nazionale di Fisica Nucleare;
the Ministry of Education, Science, Sports and Culture of Japan; 
the Natural Sciences and Engineering Research Council of Canada; 
the National Science Council of the Republic of China; 
the Swiss National Science Foundation; 
the A.~P.~Sloan Foundation; 
the Bundesministerium fuer Bildung und Forschung, Germany; 
the Korea Science  and Engineering Foundation (KoSEF), 
the Korea Research Foundation;
and the Comision Interministerial de Ciencia y Tecnologia, Spain.

%
%

\def\Journal#1#2#3#4{{#1} {\bf #2}, #3 (#4)}
\def\NCA{Nuovo Cimento}
\def\NIM{Nucl. Instrum. Methods}
\def\NIMA{{Nucl. Instrum. Methods} A}
\def\NP{Nucl. Phys.} 
\def\NPB{{Nucl. Phys.} B}
\def\PLB{{Phys. Lett.}  B}
\def\PRL{Phys. Rev. Lett.}
\def\RPP{Rep. Prog. Phys.}
\def\PRD{{Phys. Rev.} D}
\def\PR{Phys. Rep.}
\def\PRP{Prog. Theor. Phys}
\def\ZPC{{Z. Phys.} C}
\def\MPL{{Mod. Phys. Lett.} A}
\def\EPJC{{Eur. Phys. J.} C}
\def\CPC{Comput. Phys. Commun.}

\renewcommand{\baselinestretch}{1}
\mytableorig

\end{document}

%% file: cdf_auth_030501.tex
\font\eightit=cmti8
\def\r#1{\ignorespaces $^{#1}$}
\hfilneg
\begin{sloppypar}
\noindent
D.~Acosta,\r {14} T.~Affolder,\r {25} H.~Akimoto,\r {51}
M.~G.~Albrow,\r {13} D.~Ambrose,\r {37}   
D.~Amidei,\r {28} K.~Anikeev,\r {27} J.~Antos,\r 1 
G.~Apollinari,\r {13} T.~Arisawa,\r {51}
% ...
\end{sloppypar}
\vskip .026in
\begin{center}
(CDF Collaboration)
\end{center}

\vskip .026in
\begin{center}
\r 1  {\eightit Institute of Physics, Academia Sinica, Taipei, Taiwan 11529, 
Republic of China} \\
\r 2  {\eightit Argonne National Laboratory, Argonne, Illinois 60439} \\
\r 3  {\eightit Istituto Nazionale di Fisica Nucleare, University of Bologna,
I-40127 Bologna, Italy} \\
\r 4  {\eightit Brandeis University, Waltham, Massachusetts 02254} \\
\r 5  {\eightit University of California at Davis, Davis, California  95616} \\
\r 6  {\eightit University of California at Los Angeles, Los 
Angeles, California  90024} \\ 
\r 7  {\eightit University of California at Santa Barbara, Santa Barbara, California 
93106} \\ 
\r 8 {\eightit Instituto de Fisica de Cantabria, CSIC-University of Cantabria, 
39005 Santander, Spain} \\
\r 9  {\eightit Carnegie Mellon University, Pittsburgh, Pennsylvania  15213} \\
\r {10} {\eightit Enrico Fermi Institute, University of Chicago, Chicago, 
Illinois 60637} \\
\r {11}  {\eightit Joint Institute for Nuclear Research, RU-141980 Dubna, Russia}
\\
\r {12} {\eightit Duke University, Durham, North Carolina  27708} \\
\r {13} {\eightit Fermi National Accelerator Laboratory, Batavia, Illinois 
60510} \\
\r {14} {\eightit University of Florida, Gainesville, Florida  32611} \\
\r {15} {\eightit Laboratori Nazionali di Frascati, Istituto Nazionale di Fisica
               Nucleare, I-00044 Frascati, Italy} \\
\r {16} {\eightit University of Geneva, CH-1211 Geneva 4, Switzerland} \\
\r {17} {\eightit Glasgow University, Glasgow G12 8QQ, United Kingdom}\\
\r {18} {\eightit Harvard University, Cambridge, Massachusetts 02138} \\
\r {19} {\eightit Hiroshima University, Higashi-Hiroshima 724, Japan} \\
\r {20} {\eightit University of Illinois, Urbana, Illinois 61801} \\
\r {21} {\eightit The Johns Hopkins University, Baltimore, Maryland 21218} \\
\r {22} {\eightit Institut f\"{u}r Experimentelle Kernphysik, 
Universit\"{a}t Karlsruhe, 76128 Karlsruhe, Germany} \\
\r {23} {\eightit High Energy Accelerator Research Organization (KEK), Tsukuba, 
Ibaraki 305, Japan} \\
\r {24} {\eightit Center for High Energy Physics: Kyungpook National
University, Taegu 702-701; Seoul National University, Seoul 151-742; and
SungKyunKwan University, Suwon 440-746; Korea} \\
\r {25} {\eightit Ernest Orlando Lawrence Berkeley National Laboratory, 
Berkeley, California 94720} \\
\r {26} {\eightit University College London, London WC1E 6BT, United Kingdom} \\
\r {27} {\eightit Massachusetts Institute of Technology, Cambridge,
Massachusetts  02139} \\   
\r {28} {\eightit University of Michigan, Ann Arbor, Michigan 48109} \\
\r {29} {\eightit Michigan State University, East Lansing, Michigan  48824} \\
\r {30} {\eightit Institution for Theoretical and Experimental Physics, ITEP,
Moscow 117259, Russia} \\
\r {31} {\eightit University of New Mexico, Albuquerque, New Mexico 87131} \\
\r {32} {\eightit Northwestern University, Evanston, Illinois  60208} \\
\r {33} {\eightit The Ohio State University, Columbus, Ohio  43210} \\
\r {34} {\eightit Osaka City University, Osaka 588, Japan} \\
\r {35} {\eightit University of Oxford, Oxford OX1 3RH, United Kingdom} \\
\r {36} {\eightit Universita di Padova, Istituto Nazionale di Fisica 
          Nucleare, Sezione di Padova, I-35131 Padova, Italy} \\
\r {37} {\eightit University of Pennsylvania, Philadelphia, 
        Pennsylvania 19104} \\   
\r {38} {\eightit Istituto Nazionale di Fisica Nucleare, University and Scuola
               Normale Superiore of Pisa, I-56100 Pisa, Italy} \\
\r {39} {\eightit University of Pittsburgh, Pittsburgh, Pennsylvania 15260} \\
\r {40} {\eightit Purdue University, West Lafayette, Indiana 47907} \\
\r {41} {\eightit University of Rochester, Rochester, New York 14627} \\
\r {42} {\eightit Rockefeller University, New York, New York 10021} \\
\r {43} {\eightit Instituto Nazionale de Fisica Nucleare, Sezione di Roma,
University di Roma I, ``La Sapienza," I-00185 Roma, Italy}\\
\r {44} {\eightit Rutgers University, Piscataway, New Jersey 08855} \\
\r {45} {\eightit Texas A\&M University, College Station, Texas 77843} \\
\r {46} {\eightit Texas Tech University, Lubbock, Texas 79409} \\
\r {47} {\eightit Institute of Particle Physics, University of Toronto, Toronto
M5S 1A7, Canada} \\
\r {48} {\eightit Istituto Nazionale di Fisica Nucleare, University of Trieste/\
Udine, Italy} \\
\r {49} {\eightit University of Tsukuba, Tsukuba, Ibaraki 305, Japan} \\
\r {50} {\eightit Tufts University, Medford, Massachusetts 02155} \\
\r {51} {\eightit Waseda University, Tokyo 169, Japan} \\
\r {52} {\eightit University of Wisconsin, Madison, Wisconsin 53706} \\
\r {53} {\eightit Yale University, New Haven, Connecticut 06520} \\
\end{center}

%% file: prl_05072003.bbl
\begin{thebibliography}{99}
\bibitem{cdfdefs}
 We use a coordinate system where $\theta$ and $\phi$ are the polar
  and azimuthal angles, respectively, with respect to the proton
  beam direction ($z$ axis). The pseudorapidity $\eta$ is defined as
  $-\ln [ \tan(\theta/2) ]$.  
  The transverse momentum of a particle is denoted as
  $\pt = p \sin\theta$.  
  The analogous quantity using energies, 
  defined as $E_T = E \sin\theta$,
  is called transverse energy. 
  The missing transverse energy, $\met$, is a magnitude of
  $\metvec \equiv  -\sum E_T^i \,\hat{n}_i$, 
  where $\hat{n}_i$ is the unit vector in the
  transverse plane pointing from the interaction point to
  the energy deposition in calorimeter cell~$i$.
\bibitem{ETDef}  $\pttauh$  is  defined as the sum of the $\pt$ of any tracks in a 10\degr\ cone around the center of the candidate, plus the \et\ of any identified $\pi^0$'s, as measured in the EM calorimeter.
\bibitem{Zuncert} For the electron channel we have $N_{\Zz}^{Obs}- N_{\Zz}^{BG} = 54-(8.1\pm$2.5) events which gives 46$\pm$8 events when statistical uncertainties are taken into account. Similarly, for the muon channel we have $23-(2.9\pm$1.5) which gives 20$\pm$5 events.
\bibitem{IndivResults} We note for completeness that the estimated systematic uncertainties in $R_{Acc}$ due to $\stoppair$ production and decay for the stop mass range from 130 to 70~\gevcc\ are: between 4.5 and 8.2\% due to choice of the  $Q^2$ scale (taken to be correlated, and equal for the $e$ and $\mu$ cases), 2.0 and 4.6\%  due to the choice in PDFs (again taken to be correlated and equal for $e$ and $\mu$), 2.3 and 6.4\%  due to uncertainty in the initial and final state gluon radiation (correlated, and averaged between $e$ and $\mu$), 1.1 and 3.7\% due to jet energy scale (correlated and averaged), and 1.7 and 4.7\% for $e$'s and 2.3 and 4.8\% for $\mu$'s due to MC statistics (uncorrelated).
\end{thebibliography}
